\documentclass[structabstract]{aa}  
\usepackage{natbib}
\usepackage{aalongtable}
\bibpunct{(}{)}{;}{a}{}{,} 
\usepackage{graphicx}
\usepackage{txfonts}
\newcommand{\hi}{{H\sc{i}~}}
\begin{document}
   \title{The four leading arms of the Magellanic Cloud system}

   \subtitle{Evidence for interaction with Milky Way disk and halo}

   \author{M.S. Venzmer
          \inst{1,2}, J. Kerp \inst{1}, P.M.W. Kalberla \inst{1}
          }

   \institute{$^1$Argelander-Institut f\"ur Astronomie,
              Auf dem H\"ugel 71, D-53121 Bonn, Germany\\
$^2$Institut f\"ur Astrophysik, Friedrich-Hund-Platz 1, D-37077 G\"ottingen, Germany\\
              }

   \date{Received December 19, 2011; accepted August 13, 2012}

 
  \abstract
  {The Magellanic Cloud System (MCS) interacts via tidal and drag forces with the Milky Way galaxy. } 
  {Using the Parkes Galactic All--Sky Survey (GASS) of atomic hydrogen we explore the role of drag on the evolution of the so-called Leading Arm (LA). }
  {We present a new image recognition algorithm that allows us to differentiate features within a 3-D data cube (longitude, latitude, radial velocity) and to parameterize individual coherent structures. We compiled an \hi object catalog of LA objects within an area of 70\degr x 85\degr (1.6 sr) of the LA region. 
This catalog\thanks{Table 2 is only available in electronic form at the CDS via anonymous ftp to cdsarc.u-strasbg.fr (130.79.128.5) or via http://cdsweb.u-strasbg.fr/cgi-bonn/qcat?J/A+A/} comprises information of location, column density, line width, shape and asymmetries of the individual LA objects above the 4--$\sigma$ threshold of $\Delta T_b \simeq 200\,{\rm mK}$. }
  {We present evidence of a fourth arm segment (LA4). For all LA objects we find an inverse correlation of velocities $v_{\rm GSR}$ in Galactic Standard of Rest frame with Magellanic longitude. High--mass objects tend to have higher radial velocities than low--mass ones. About 1/4 of all LA objects can be characterized as head-tail (HT) structures. 
    }
  {Using image recognition with objective criteria, it is feasible to isolate most of  LA emission from the diffuse Milky Way \hi gas. Some blended gas components (we estimate 5\%) escape detection, but we find a total gas content of the LA that is about 50\% higher than previously assumed. These methods allow the deceleration of the LA clouds to be traced towards the Milky Way disk by drag forces. The derived velocity gradient strongly supports the assumption that the whole LA originates entirely in the Large Magellanic Cloud (LMC). LA4 is observed opposite to LA1, and we propose that both arms are related, spanning about 52\,kpc in space. HT structures trace drag forces even at tens of kpc altitudes above the Milky Way disk.}
 
\titlerunning{The 4 Leading Arms of the MCS}

\maketitle
%

\section{Introduction}
Because of its proximity to the Milky Way, the MCS forms an ideal laboratory for studing the structure formation within the local universe in great detail. The more massive Milky Way galaxy accrete the MCS and tidal forces and perhaps drag separate gas from the stellar bodies. High--sensitivity optical studies disclose even the faintest stellar populations of evolved stars and allow to search for evidence of the stellar--gas--feedback processes triggered by the interaction with the Milky Way halo and/or its gravitational field \citep{indu11}. Both stellar distributions are highly concentrated, so only a fraction also populate the Magellanic Bridge \citep{putman03, irwin1990}. So far the prominent \hi structures denoted as the Leading Arm \citep[LA,][]{putman1998} and the Magellanic Stream have not detectable stellar counterpart \citep[MS,][]{wannier1972, putman03, bruens05}. 

Several investigations of the MS make use of specialized surveys \citep{bruens05} or the Leiden-Argentine-Bonn all-sky survey \citep{nidever08}. Even though a wealth of information has been compiled today \citep[see][and references therein for a recent review]{nidever08}, it is still a matter of debate to what extent tidal or drag forces determine the structure formation of the MS and LA. Not detecting of stars in the MS argues for drag forces, decelerating only the viscous gaseous component, while the stellar component is unaffected and continues its orbital motion.  The existence of the LA, however still favors strongly a tidal origin \citep{nidever08}.

In this paper we focus on the \hi 21-cm emission distribution of the LA as observed with the Galactic All-Sky Survey \citep[GASS,][]{naomi09,kalb10}.
Assuming a linear distance of about 50\,kpc from the Sun \citep{cioni00}, single--dish radio telescopes like the Parkes 64-m dish offer a linear resolution of about 200\,pc at \hi 21-cm wavelength.
The high spectral resolution of state-of-the-art radio spectrometers not only allows determination of the bulk motion of individual clouds but also examination of well resolved \hi line profiles.
GASS for the first time makes fully sampled sensitive wide field images of the \hi distribution available.

For our study we extracted a huge portion of the southern sky from this data base to compile a comprehensive catalog of LA clouds and filaments (Sect.\,\ref{sec:data}).
To achieve this aim we developed an approach to differentiating between Milky Way gas and superposed (in space and frequency) MCS \hi emission. Based on a newly developed extension of a standard image recognition algorithm, we decompose the \hi data without the necessity of Gaussian fits to the emission line profiles (Sect.\,\ref{sec:decompose}).
This allows us to compile a comprehensive view of the LA down to a limit $T_{\rm B} = 200$\,mK. 
Using this catalog we investigate the velocity structure of the LA and a newly identified leading arm four feature as a whole (Sect.\,\ref{sec:velostruc}).
The catalog comprises the basic physical parameters of all identified LA objects; this gives us the opportunity to analyze the occurrence of head-tail structures and their orientation with respect to the MCS (Sect.\,\ref{sec:HT}).
We end the paper with a brief summary (Sect.\,\ref{sec:summary}).


\section{The data}
\label{sec:data}
We use GASS data, observed with the L-band multi-feed receiver at the Parkes
64m telescope. Observations and initial data processing of the survey were
described by \citet{naomi09}. We extracted data from the final data release
\citep{kalb10} that were corrected for stray radiation, instrumental effects,
and radio-frequency interference (RFI). In calculating FITS cubes we used the web service as provided by
$http://www.astro.uni-bonn.de/hisurvey/$ but with two major extensions: 1) the
positions were converted to the Magellanic Stream coordinate system as defined
by \citet{nidever08}, 2) The radial velocities were transformed into the
galactic-standard-of-rest system (GSR) \citep[Eq. 2 of][]{deHeij02} to 
quantify the motion of the MCS relative to the Milky Way galaxy. We consider the GSR system as best suited to evaluating the dynamics of the coherent MCS structures relative to the Milky Way galaxy. This is most useful for visualization; however, it does not allow ambiguities to be overcome when LA clouds are blended with Milky Way gas.  

The GASS web server allows generating maps with different predefined angular
resolutions \citep{kalb10}.
 For our data analysis we chose two settings with effective beams of FWHM = 14.4\,arcmin (4.8\,arcmin per data cube pixel) and FWHM = 20.7\,arcmin, resulting in average rms uncertainties of 50\,mK and 25\,mK, respectively, for channel maps at a spectral resolution of
The smoothed data were used to initialize our source finding algorithm (Sect. \ref{sec:imagej and source identification}), all further processing was done on the high resolution data base.

The whole field of interest covers $0^\circ \leq l_{\rm LMC} \leq 70^\circ$
and $-42.5^\circ \leq b_{\rm LMC} \leq 42.5^\circ$. This corresponds to
1.59\,sr or 13\% of the whole sky (Fig.\,\ref{LA_clouds} left). The radial
velocity interval of interest covers $-80\,{\rm km\,s^{-1}} \leq v_{\rm GSR}
\leq 250\,{\rm km\,s^{-1}}$. According to \citet{vandermarel02}, the LMC
velocity is $v_{\rm GSR} = 84\,\pm 7\,{\rm km\,s^{-1}}$. Inspecting
Fig.\,\ref{posvel} (top panel) shows that the negative velocity limit cannot be
determined very well because of the confusion with the Milky Way galaxy
emission. The positive velocity limit safely includes all MCS
features above our 4--$\sigma$ threshold of $T_{\rm B} = 200$\,mK (Fig.\,\ref{LA_clouds}
right).


\section{Decomposing the LA emission}
\label{sec:decompose}

Our aim was to compile an object catalog for the LA area, containing
all data that are necessary for any interpretation, such as position,
velocity, brightness temperature, center of emission, geometrical center, column density, and velocity gradient. For an accurate parameterization we also decided to determine first moments in the position-velocity
domain. This allows us to investigate the cloud's morphology and quantify their shape, kinematics, and column density distribution asymetries in detail.
The resulting catalog comprises emission features of all kind of structures, regardless of their physical origin. 

A major task is to eliminate spurious features and to distinguish between
sources belonging to LA and the Milky Way. \citet{nidever08} use a Gaussian decomposition for this purpose, which is a very elaborate enterprise. Intuitively, a parameterization in first moments, as mentioned above, also enable us to distinguish between
different object classes according to their cataloged parameters, such as full area features (i.e. the Milky Way), ``edge'' objects (gas clouds touching the limits of the data cube either in position or velocity), small objects, and known galaxies. It is clear
that the available source parameters must be complete enough to distinguish between 
different object classes.

\subsection{ImageJ and source identification}	
\label{sec:imagej and source identification}

The cloud catalog is automatically compiled by a newly developed source finding and parametrization program within the ImageJ enviroment. ImageJ is a scientific image processing software \citep{Rasband2009} that is most commonly used in biology and medicine but also partly in astronomy. It is an open source software, and it can be found at \texttt{http://rsbweb.nih.gov/ij/index.html}.

The source--finding algorithm is designed to identify both spatially and kinetically coherent \hi objects above a given brightness temperature threshold. For each object some essential physical parameters are derived, as presented above. 

\subsubsection{Cleaning the catalog}	
\label{sec:cleaning the catalog}
First we clean our database for insignificant features below the instrumental noise, for RFI, and residual systematical baseline defects. We use the smoothed data with a doubled kernel size of $0.25^\circ$. All features below the threshold of $T_{\rm B} = 200$\,mK can be excluded by this approach. Based on this threshold, we generated a mask for the high--resolution data cube. 

The threshold mask isolates coherent structures and different individual objects. These structures need to be separated from other objects in phase space (spatially but also in velocity). This procedure attributes each pixel above the threshold to an associated object, permitting the calculation of the objects physical parameters. 

The GASS beam FWHM ($14.4$\,arcmin$ = 0.24^\circ$) corresponds to three pixels in the data cube. By this, any detected objects smaller than a single Parkes beam were excluded from further analysis. With this step we mitigate the imprint of RFI. We cross correlated our catalog with the HIPASS bight galaxy catalog \citep{Koribalski2004}. Within the area of interest, 361 galaxies are located. Most of them are automatically excluded from further consideration because their velocities are beyond the velocity range considered here.
Only three residual galaxies were needed to be accounted manually.

Cold gas with an FWHM line width of up to three spectral channels ($2.4$~km\,s$^{-1}$) are considered to trace mainly the Milky Way emission because a sufficient high gas pressure is needed to warrant cold gas of $\la 250 $K in a two-phase medium \citep{wolfire1995b}. HVCs containing extremely cold dense cores will still be cataloged because next to the narrow \hi emission the enveloping warm gas will also be detected in adjacent spectral channels. RFI events which are however, usually narrowly confined in frequency, will be rejected. 

After such a selection, the LA catalog in its final version has 449 entries. In the following we show that these cataloged objects might safely be considered as belonging to the LA because of their coherent velocity structure. Towards the field of interest, next to the LA, the Wannier complexes are well known \citep{wannier1972}. These HVC complexes (annotated in Fig.\,\ref{posvel} and discussed in Sect. 4) form coherent structures but are separated from most of the LA structures by their lower GSR velocities.

\subsubsection{The Milky Way galaxy as an ``edge'' object}	
Based on the catalog it is feasible to generate data cubes showing different representations of the LA objects.
The spacial distribution of all compiled objects is shown in Fig.\,\ref{LA_clouds}(left panel). It includes objects that are boundary or edge objects of the data cube. These edge objects were excluded from further consideration (see Fig.\ref{LA_clouds} right panel).
The comparison of both panels make it evident that the Milky Way disk emission is successfully removed from the analyzed catalog.
The Milky Way disk \hi emission above the threshold of 200\,mK forms a single broad and coherent structure which touches all boundaries of the whole data cube. By definition the Milky Way galaxy \hi emission is classified as an ``edge'' object. Accordingly, when dropping the edge objects from the final catalog, the Milky Way emission is entirely removed.

It is a big advantage of the applied algorithm  to distinguish easily between the LA and the Milky Way disk without the need to fit other parameters.
However, \hi objects physically belonging to the MCS but apparently ``connected'' with the \hi emission of the Milky Way galaxy will not be part of the final catalog. Here visual inspection might further improve the completeness of the catalog but could also introduce systematic biases. 

\subsection{The cataloged parameters}	
\label{sec:the cataloged parameters}

An extensive set of parameters is calculated for each cataloged object. The basic parameters are the position, the spatial and the kinematical centers, the intensity weighted center, the geometrical center, and the  position of maximum brightness temperature. Further compiled parameters are the object extent in all three cube dimensions, the maximum brightness temperature, the maximum column density, the mean column density, the integrated brightness temperature, the phase space volume, the solid angle, the orientation angle, and the FWHM velocity width. The kinetic gas temperature is derived from the full-width-half maximum of the \hi line of the peak brightness temperature position. Assuming a $50$~kpc distance of the LA, the linear extent and the mass are calculated as well (see Tab.\,\ref{tab:LA_Objects}).

\begin{figure*}
   \centering
   \includegraphics[width=13cm, angle=270]{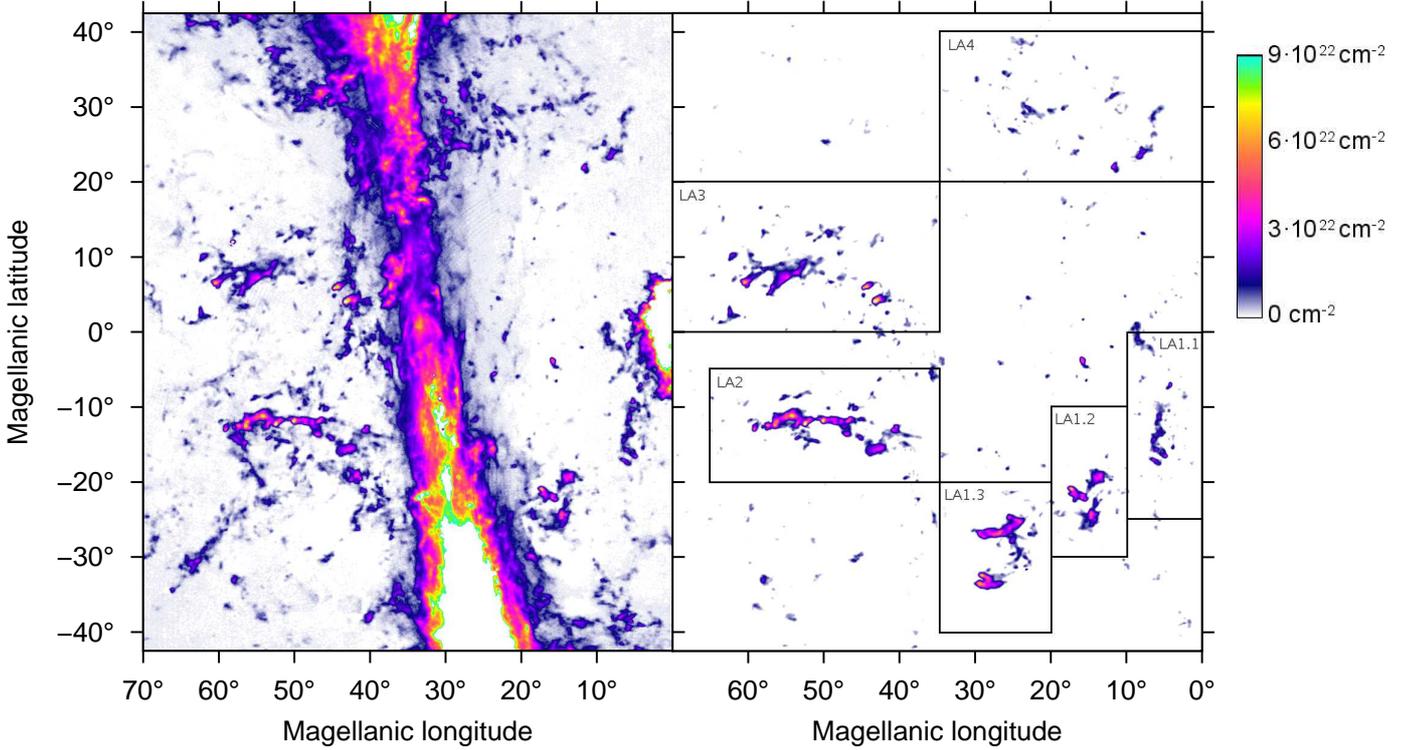}
   \caption{{\bf Left:} The map shows the column density distribution of the \hi 21-cm line emission across the velocity interval $-80\,{\rm km\,s^{-1}} \leq v_{\rm GSR} \leq 250\,{\rm km\,s^{-1}}$. The complex \hi structure is produced by the superposition of the Milky Way galaxy emission, the LMC, and the LA.
{\bf Right:} The column density distribution of LA clouds across $-80\,{\rm km\,s^{-1}} \leq v_{\rm GSR} \leq 250\,{\rm km\,s^{-1}}$ starting at a threshold of $\Delta T_B = 200\,{\rm mK}$ ($4-\sigma$ level of the GASS data). All the objects shown are cataloged and parametrized. Coherent structures like the Milky Way disk and LMC emission are according to the applied criteria ``border'' objects and not part of the catalog. The boxed and labeled areas enclose coherent structures in the data cube. Objects outside these areas also belong most likely to the LA. {\bf Note:} White colors inside regions with disk emission denote column densities in excess of $N_{\rm HI} = 9\cdot 10^{22}\,{\rm cm^{-2}}$.}
             \label{LA_clouds}
    \end{figure*}

\begin{figure*}
   \centering
   \includegraphics[width=9.5cm, angle=270]{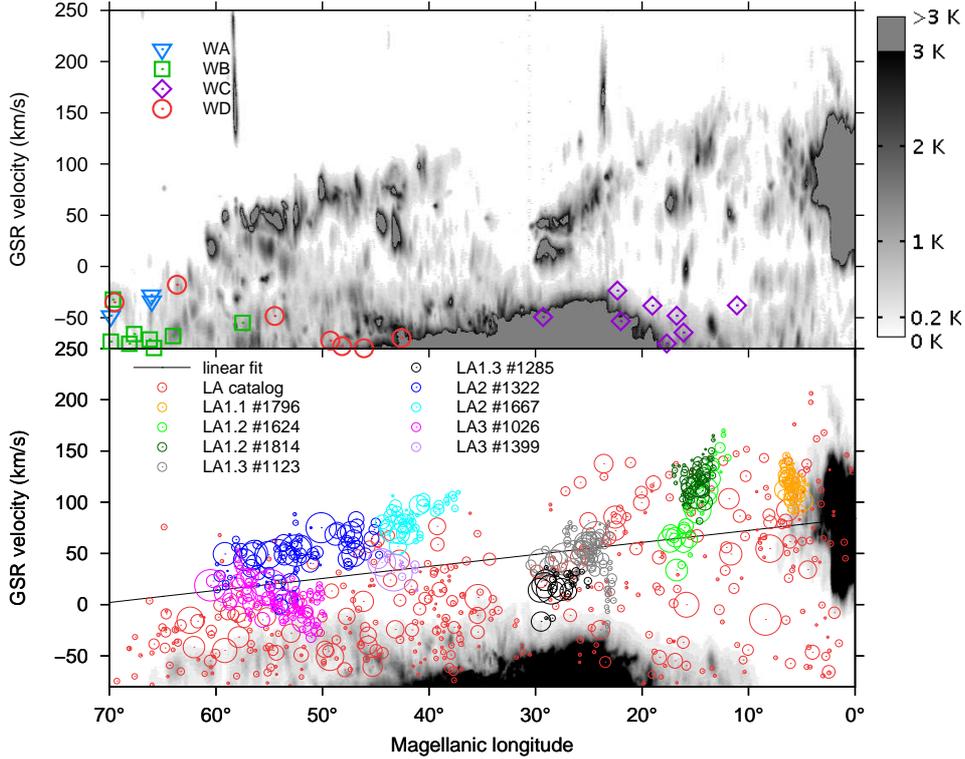}
\vspace{0.5cm}
   \caption{{\bf Top:} Position ($l_{\rm MS}$) versus velocity ($v_{\rm GSR}$) diagram showing the maximum brightness temperature distribution across $-42.5^\circ \leq b_{\rm LMC} \leq 42.5^\circ$ across the whole velocity range of interest (starting at $T_{\rm B} = 200\,{\rm mK}$ up to 3\,K on a linear scale). On the righthand side the LMC \hi emission, while the Milky Way emission is visible towards the bottom. The narrowly confined high--velocity dispersion emission lines are associated with nearby galaxies covered by GASS. Obvious is a continuous velocity gradient of the \hi emission above the Galactic plane from the LMC systemic velocity to $v_{\rm GSR} = 0\,{\rm km\,s^{-1}}$. The location of clouds belonging to the Wannier complexes WA, WB, WC, and WD are annotated. 
{\bf Bottom:} The colored circles denote the location of the individual LA catalog objects   while the gray-scale figure represents the ``edge objects'', the Milky Way and the LMC. Each color marks an individual LA cloud complex (the ID of the catalog -- available at the CDS -- entry is given), aiming to demonstrate the coherent velocity structure of the complexes. The circle diameter scales according to $d \propto M^{-\frac{1}{3}}$. The regression line marks the mass--weighted velocity gradient with its LMC value of $v_{\rm GSR} = 84.2\pm7\,{\rm km\,s^{-1}}$ demonstrating that the LA gas most likely is released from the LMC gaseous body. 
}
 \label{posvel}
 \end{figure*}

\begin{figure*}
   \centerline{
   \includegraphics[width=17cm, angle=270]{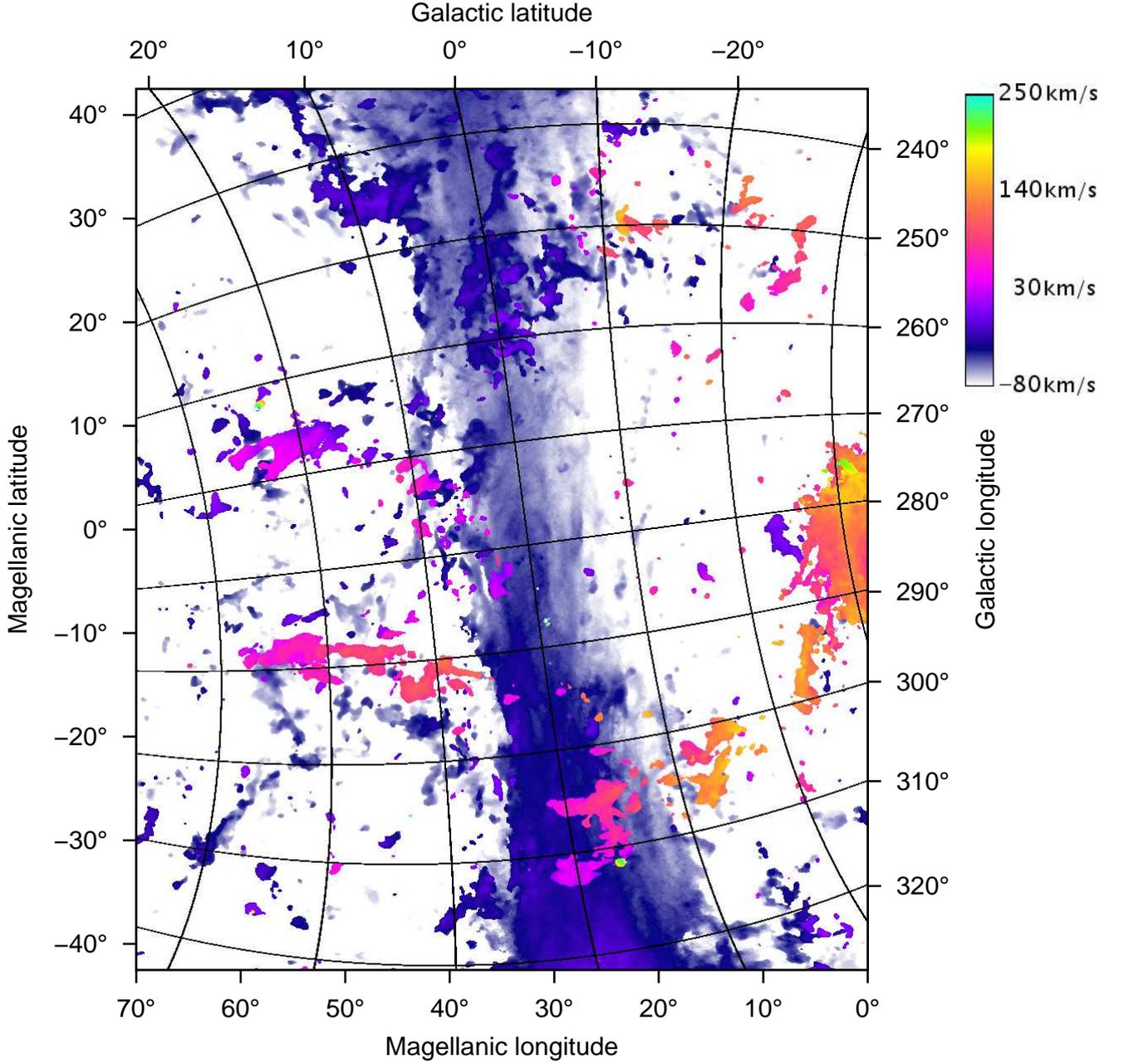}}
   \caption{This color--coded figure illustrates the applied method to differentiate between Milky Way, LMC, and LA \hi emission. The cataloged objects show up with significant different GSR velocity distributions (color), allowing them to be separated from the neighboring \hi emission within the data cube. Applying the strong criterion to separate between Milky Way and LA emission (Fig.\ref{LA_clouds} right) we might miss coherent structures, i.e. $l_{\rm MS} = 35^\circ$ and $b_{\rm MS} = 28^\circ$. These objects are continuously connected in \hi emission with the Milky Way galaxy.}
              \label{LA_total_area}%
    \end{figure*}

Figure\,\ref{LA_total_area} displays numerous features that appear well separated from the diffuse Galactic emission.
Only clouds are considered that are unambiguously separated in phase space from the Milky Way. From a comparison of Fig.\,\ref{LA_clouds} (right) with Fig.\,\ref{LA_total_area} one might expect that some fraction of the total mass assigned to the leading arm might be missing. Correspondingly we find few objects at Magellanic longitude $l_{\rm MS} \sim 33\degr$ visible in Fig.\,\ref{posvel}. We use the area filling factor for the affected region in velocity and position (the projected phase space) as a measure to estimate the mass deficit, as less than 5\%.

\subsection{Mass comparison}
\label{sec:mass comparison}

The masses of LA features derived here are about 50\% higher as recognized previously. In Table\,1 masses of LA substructures are compiled and compared with \citet{bruens05}. The obvious differences are first because of different definition of the feature extents (in general in \citet{bruens05} the features are systematically smaller). In particular LA1.1 and LA1.3 are extended objects that are partly located outside the survey area of \citet{bruens05}.

\begin{table}
	\centering
	\caption{Mass comparison of the specific LA areas indicated in Fig.\,\ref{LA_clouds} with \citet{bruens05}.}
	\begin{tabular}{lccc}
		\hline\hline
		{Region}	&$M_{\rm HI}$	&$M_{\rm HI}$(B05)	&{$\frac{M_{\rm HI}}{M_{\rm HI}({B05})}$}\\
			&(${\rm 10^{6}\,M_\odot}$)	&(${\rm 10^{6}\,M_\odot}$)	&\\
		\hline
		LA1.1 *	&2.67	&0.99	&2.69\\
		LA1.2	&4.02	&3.14	&1.28\\
		LA1.3 *	&6.38	&3.64	&1.75\\
		\hline
		LA1 *	&13.07	&8.26	&1.58\\
		LA2	&11.42	&9.09	&1.26\\
		LA3	&8.89	&7.44	&1.19\\
		LA4	&2.44	&--	&--\\
		Rest	&2.34	&--	&--\\
		\hline
		LA total *	&38.15	&24.79	&1.54\\
		\hline
	\end{tabular}
	\label{tab:masstable}
	\tablefoot{Masses from \citet{bruens05} are indicated by B05. All masses are calculated assuming a linear distance of $50$\,kpc. The asterisk (*) mark areas that are obviously not congruent; i.e., the studied areas by \citet{bruens05} are in general of smaller extent.}	
\end{table}

   \begin{figure*}
   \centering
   \includegraphics[width=12cm, angle=270]{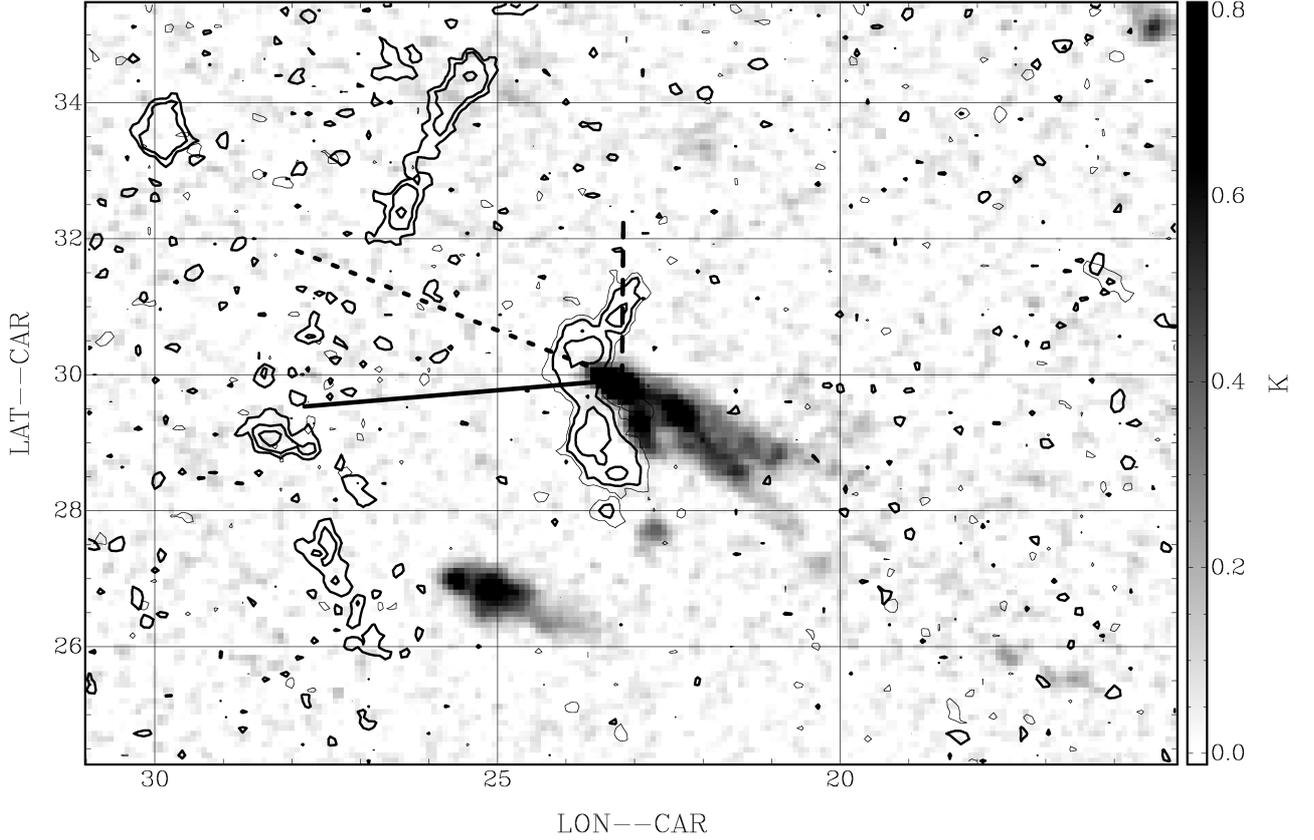}
   \caption{\hi brightness temperature distribution of two HT clouds associated with the LA4 are displayed in gray-scale on a linear scale. Both clouds show a cometary shape in this velocity channel of 84\,${\rm km\,s^{-1}}$. The contour lines mark \hi gas line temperatures at 100, 200, and 400\,mK. The closest bow--like object connected with the head of one HT cloud is at $v_{\rm GSR} = 144\,{\rm km\,s^{-1}}$, and all the more distant structures are at $v_{\rm GSR} = 25\,{\rm km\,s^{-1}}$.
The solid line marks the orbital motion of the MCS, the long dashed line the lagging rotating halo, and the short dashed line the projection of the resulting velocity vector.}
              \label{comet1}
    \end{figure*}
   \begin{figure*}
   \centerline{
   \includegraphics[width=4cm,angle=270]{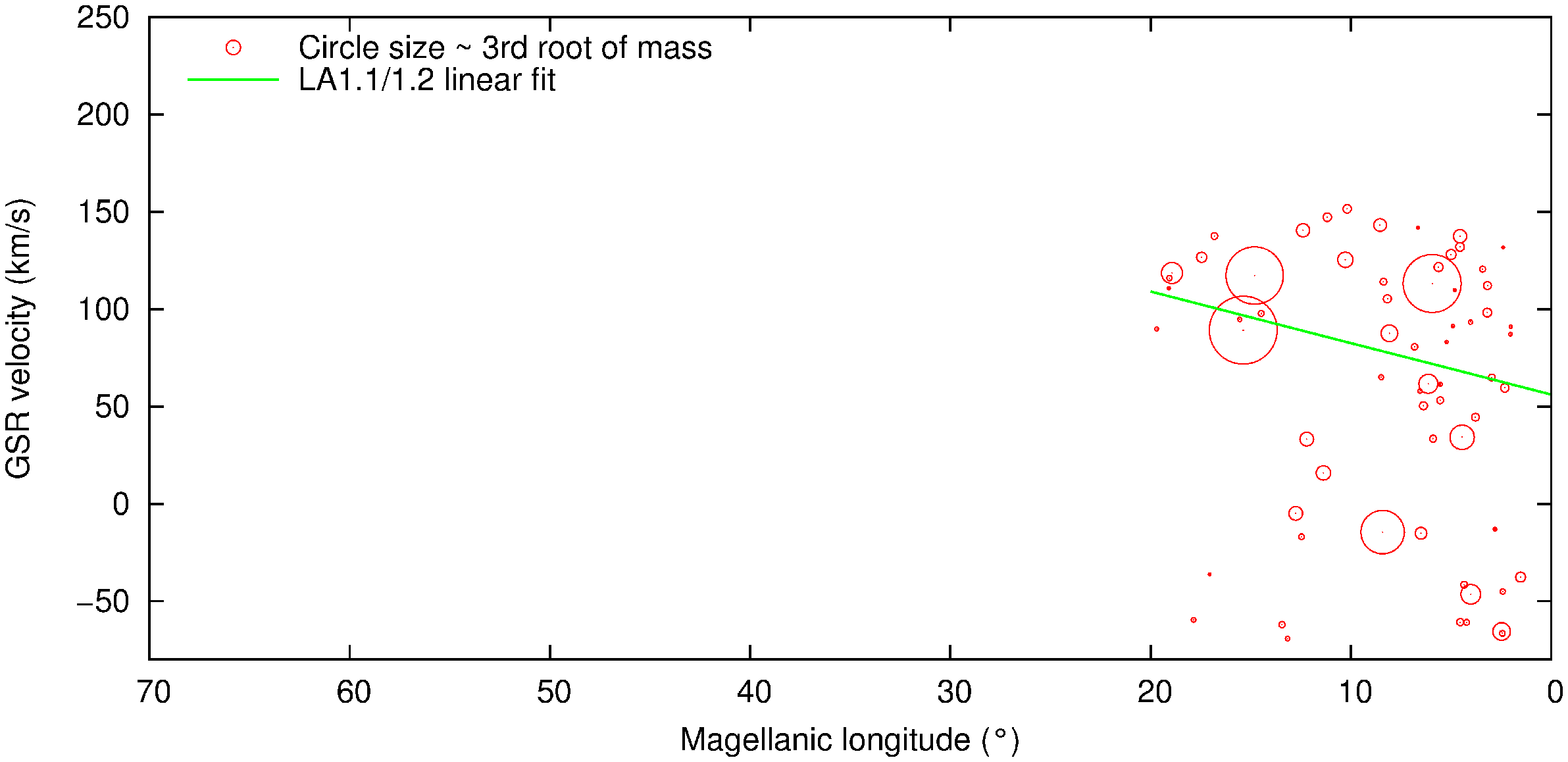}
   \includegraphics[width=4cm,angle=270]{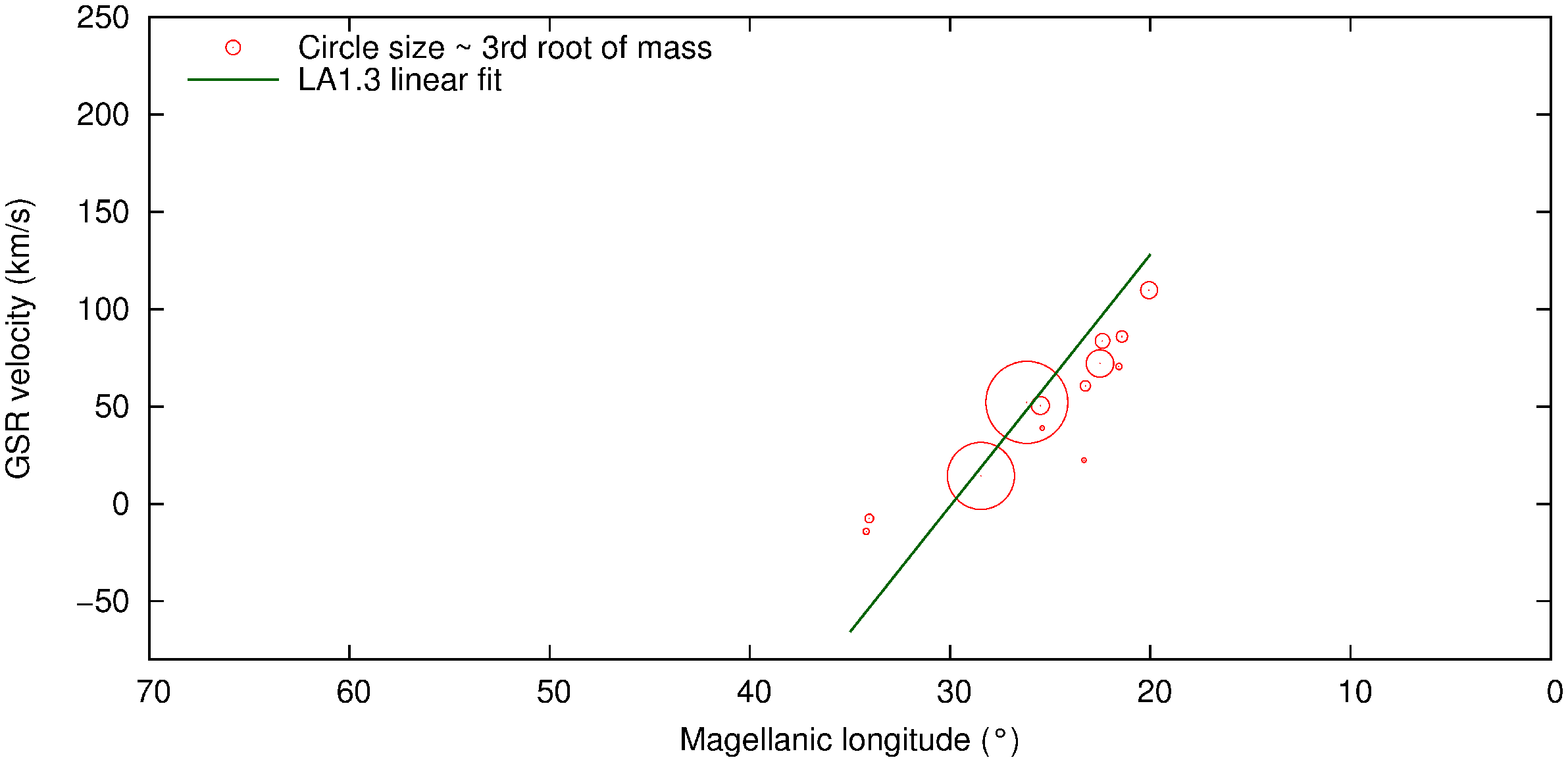}
}
   \centerline{
   \includegraphics[width=4cm,angle=270]{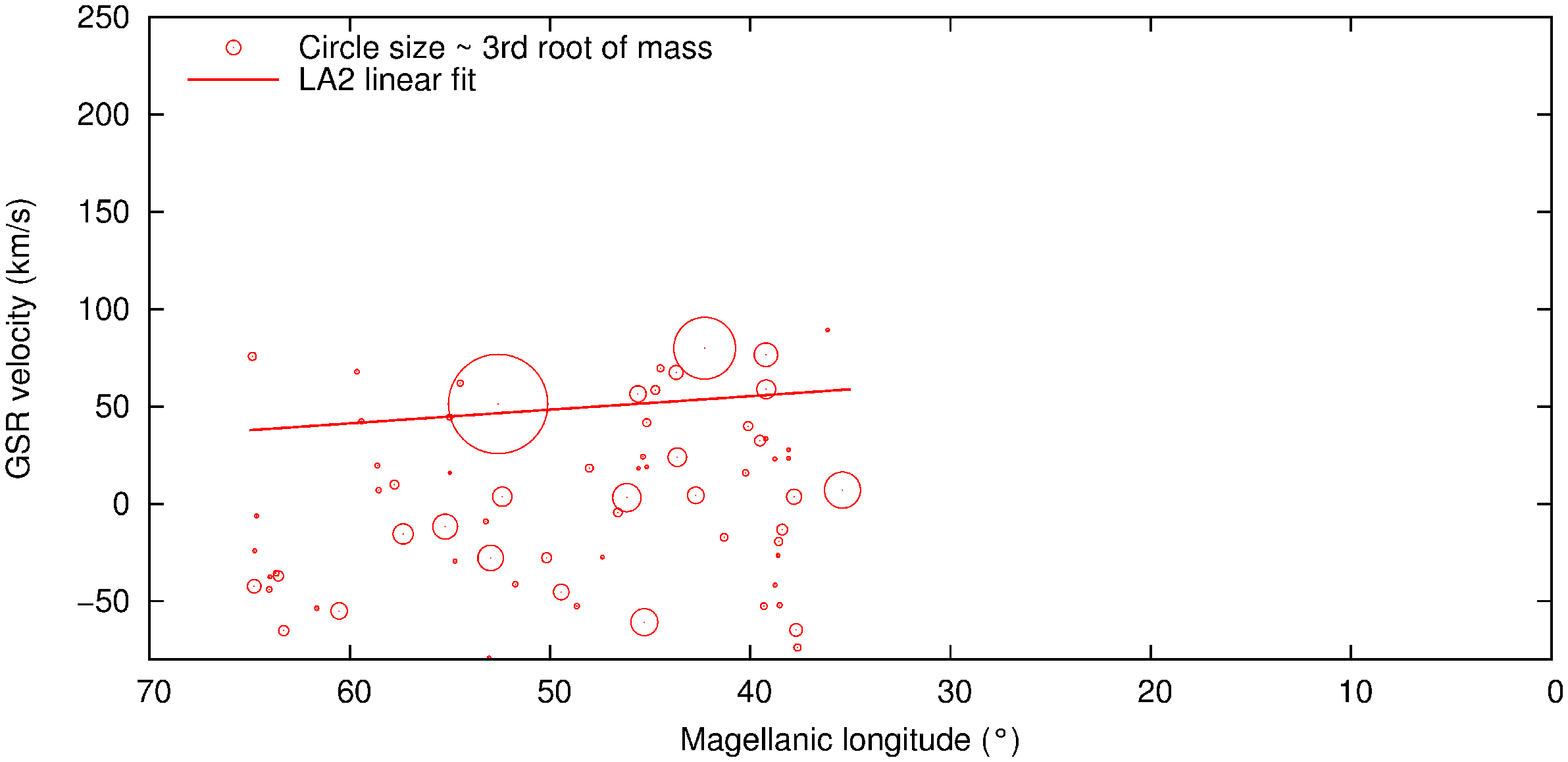}
   \includegraphics[width=4cm,angle=270]{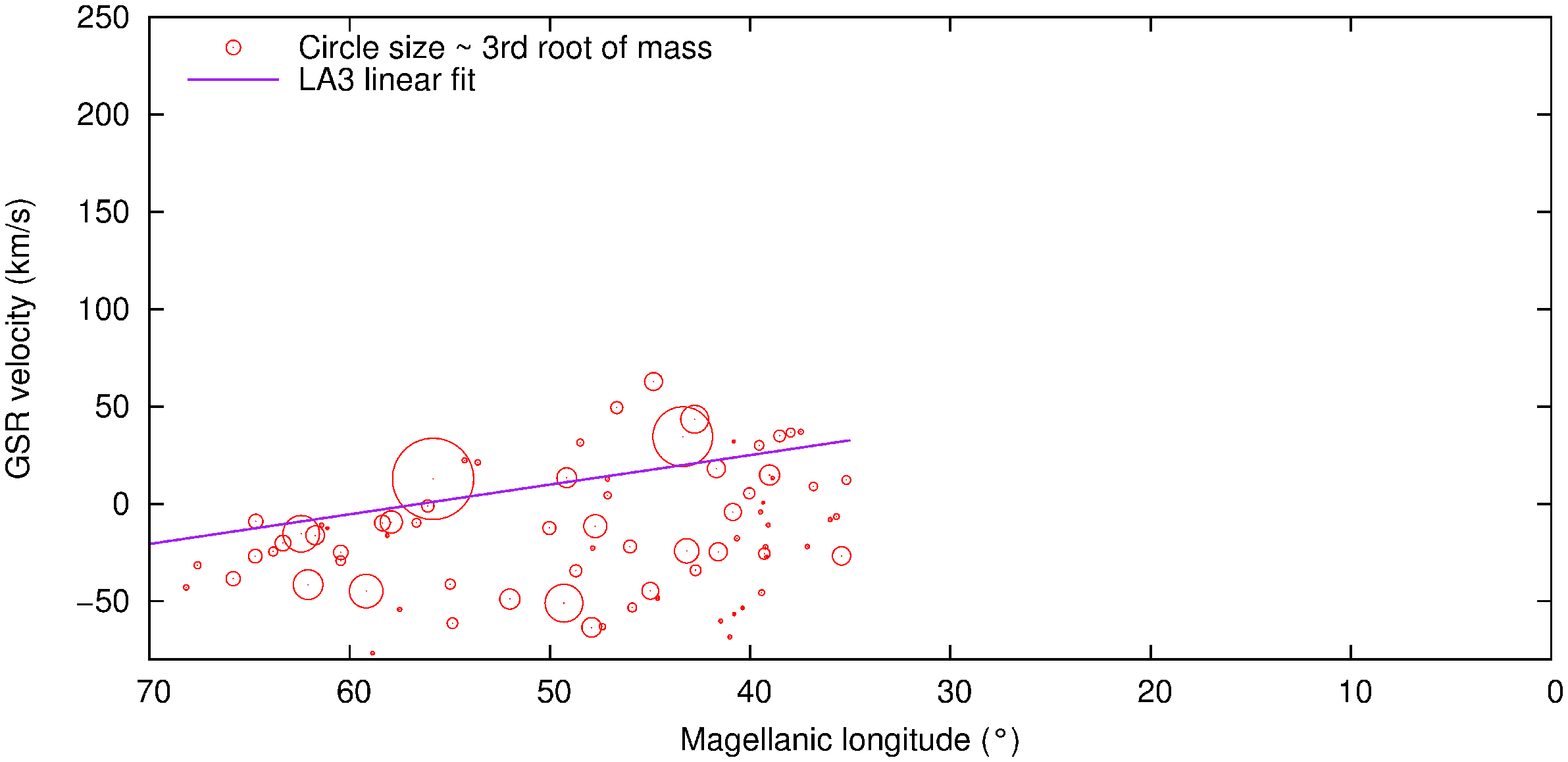}
}
   \centerline{
   \includegraphics[width=4cm,angle=270]{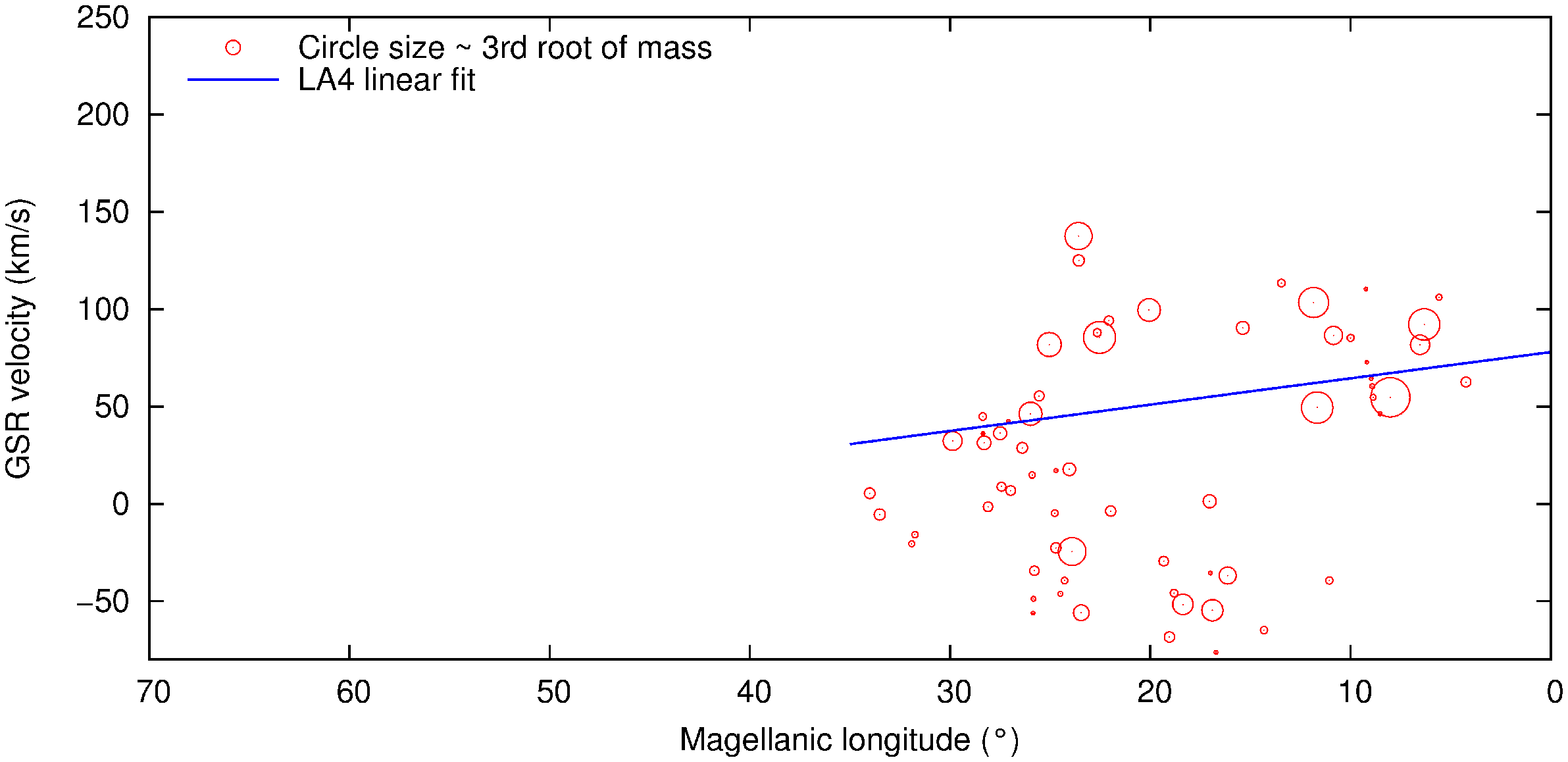}
   \includegraphics[width=4cm,angle=270]{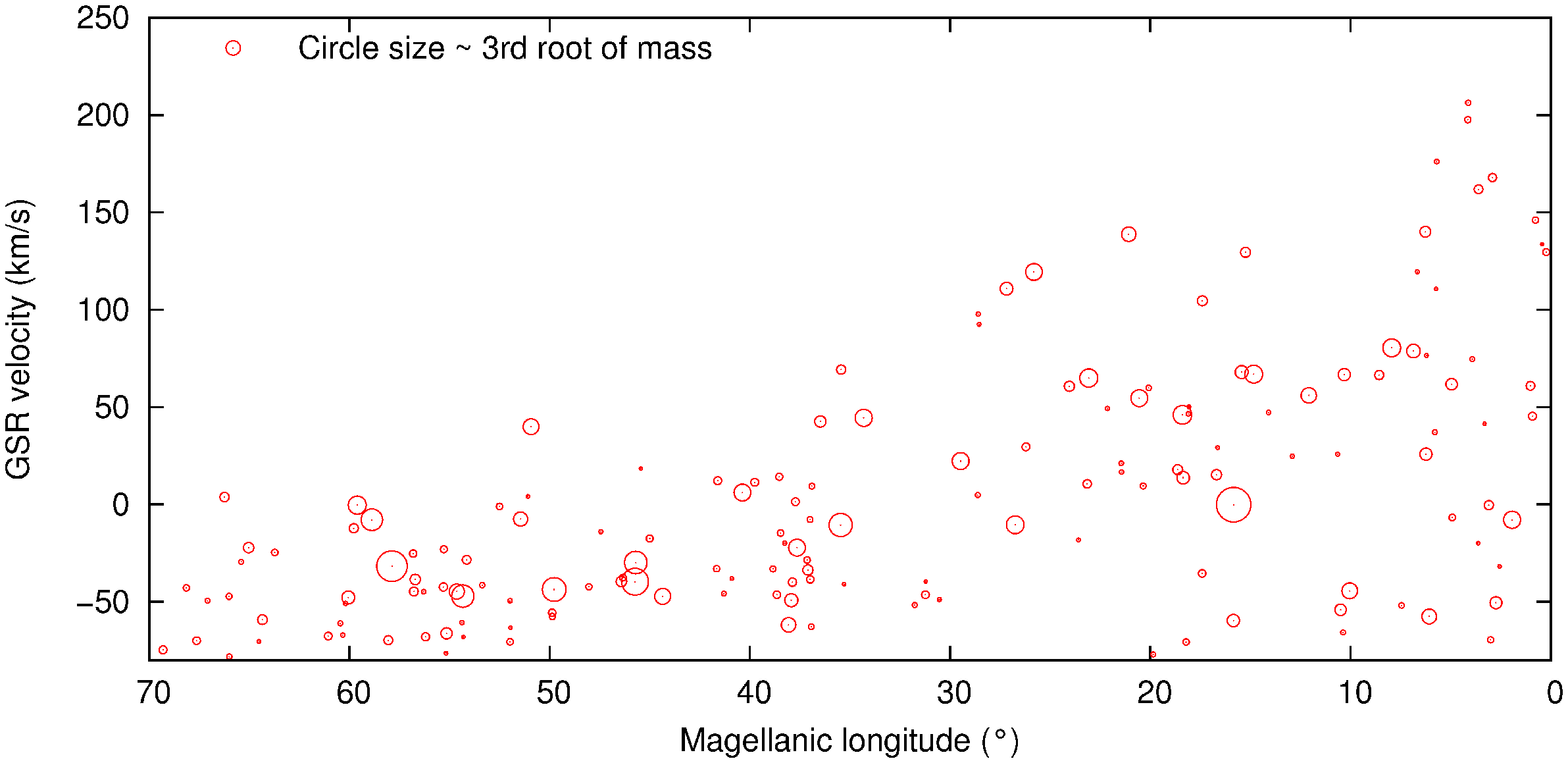}
}
   \caption{Velocity gradients of the individual coherent structures (see Fig.\ref{LA_clouds}). {\bf Top left:} LA1.1 and LA1.2 are located close to the Magellanic Clouds. Their ensemble properties are unique, and they dislose a rising $v_{\rm GSR}$ velocity with increasing $l_{\rm MS}$. {\bf Top right:} LA1.3 shows the strongest velocity gradient, consistent with the impact hypothesis proposed by \citet{naomi08}.  {\bf Middle panel:} LA2 (left panel) and LA3 (right panel) are localized at large Magellanic stream longitudes.
{\bf Bottom left:} The newly classified feature LA4 shows a velocity gradient comparable to LA2 and LA3 despite its close location to the Magellanic Clouds. {\bf Bottom right:} Those cataloged objects not accounted to the coherent structures defined in Fig.\,\ref{LA_clouds} disclose the same overall velocity gradient implying their association with the LA.}
              \label{LA_individual}
    \end{figure*}
\section{The velocity structure of the LA as a whole}\label{sec:velostruc}
We first study the velocity distribution of all LA objects together to investigate all the environmental and orbital parameters. To this aim we analyzed the radial velocity pattern of the cataloged objects in the GSR rest frame. Figure\,\ref{LA_total_area} gives an overview, while Fig.\,\ref{posvel} shows details about the individual LA objects listed in the Appendix. To distinguish between LA features and Wannier complexes, we included positions of Wannier clouds \citep{wannier1972} in the upper panel of Fig.\,\ref{posvel} as given in the \cite{WvW1991} catalog. For $l_{\rm MS} \ga 60\degr$ WA and WB features cannot be distinguished easily from scattered LA features. LA2 and LA3, however, as is obvious from  Fig.\,\ref{LA_total_area}, stand out clearly. The complexes WC and WD are found at low GSR velocities, so separated well from LA clouds.  

The diameters of the circles in the lower panel are proportional to the third root of the cloud masses and specific LA substructures are colored. All together there is an obvious relation between $l_{\rm MS}$ and $v_{\rm GSR}$. This finding implies that the cataloged LA as a whole has a coherent and continuous velocity structure. The clouds that are the most distant from the LMC tend to have the lowest velocities. Gas within the immediate vicinity of the LMC apparently had less time to get decelerated by the halo medium.

At the location of the Galactic disk (around $l_{\rm MS} = 30^\circ$, Fig.\,\ref{posvel}\,bottom panel), there is a V-shaped underabundance of LA objects. 
We attribute this under-abundance to limitations of our approach to separate Milky Way Galaxy and LA at these positions. HI clouds in this region have been attributed to Galactic emission.

Connecting the location of the cataloged LA objects with their \hi mass (assuming $D = 50$\,kpc) discloses a remarkable trend (in particular for LA\,2 and LA\,3): the higher the mass of the objects, the higher their radial velocities (Fig.\,\ref{posvel}, bottom panel and Fig.\,\ref{LA_individual} middle panel). This argues for drag which decelerates the low--mass diffuse objects significantly while the high mass objects keep most of their momentum. 
In the immediate vicinity of the LMC ($l_{\rm MS} \leq  25^\circ$, Fig.\,\ref{LA_individual} top and bottom left panels), the higher mass objects do not behave any differently in velocity.
Here, the low--mass objects are also observed at high $v_{\rm GSR}$ radial velocities.
Starting around $l_{\rm MS} \geq  35^\circ$, the higher mass objects reveal that the highest radial velocites and lower mass objects are all at lower velocities.

Performing a mass--weighted, least square approximation of the $v_{\rm GSR} (l_{\rm MS})$, we find the relation accounting for all LA objects:
\begin{equation}\label{velgrad} 
v_{\rm GSR}(l_{\rm MS}) = 84.2\,{\rm km\,s^{-1}} - 1.2\,{\rm km\,s^{-1}}\cdot l_{\rm MS\,.}
\end{equation} 
The linear regression line is shown in Fig. \ref{posvel} (bottom panel). Despite the high number of low--mass objects, the high--mass objects dominate the linear approximation of the overall velocity distribution of the LA.
Considering the global motion of the LA, the velocity at the origin of the LA, with $v_{\rm GSR} (0) = 84.2\, {\rm km\,s^{-1}}$ is consistent with the
value  $v_{\rm GSR} = 84\,\pm\,7\,{\rm km\,s^{-1}}$ as derived by \citet{vandermarel02} from an ensemble of carbon stars of the LMC.
This overall velocity structure of the \hi gas suggests that the LA has its origin in the LMC rather than in the SMC \citep{putman1998}. This is consistent with the morphological and positional coincidences of the LA1 feature with the LMC, as suggested by \citet{nidever08}, who have used quite different methods for their analysis. 

Quantitatively comparing the velocity structure of the gaseous objects of the LA with the N-body simulation of \citet{connors06} (their Fig.\,9) -- who attribute the LA to the SMC -- shows that the simulated velocity distribution is in excess of about  $200\,{\rm km\,s^{-1}}$ from the observed one. Their LMC orbit is, however, consistent with the \hi velocity structure of the high--mass objects.

\subsection{Velocity structure of LA1}\label{sec:LA1}
We distinguished three LA1 subpopulations as shown in Fig.\,\ref{LA_clouds} (right panel). Towards a single filament of LA1.3 \citet{naomi08} propose that this filament interacts with the Milky Way galaxy disk. 
\hi mass--weighted linear fits the individual LA filaments underline the general trend of decreasing $v_{\rm GSR}$ with increasing $l_{\rm MS}$. As is obvious in Fig.\,\ref{LA_individual} (top right panel) LA1.3 shows the strongest velocity gradient of all LA filaments.
Based on our statistical analysis of the LA1.3, the filament as a whole appears to be strongly decelerated consistent with \citet{naomi08}.

Considering the overall velocity gradient of the LA as an ensemble of objects (Fig.\,\ref{posvel}) and the large angular separation between the LA features (Fig.\,\ref{LA_clouds}), we propose that the LA is most likely a huge coherent structure. LA1.3 appears to pass through sufficiently dense gas of the Milky Way galaxy. LA1.2 and LA1.1 are apparently located at larger z-distances from the Galactic plane. The same is true for LA2 and LA3, but now on the opposite side of the Galactic disk, as evident from Fig.\,\ref{LA_total_area}. 

These filaments must have already passed the equator of the Galactic coordinate system, but we do not know the galactocentric distance.
In the case of an exponential radial density distribution of the Galactic disk \citep{kalberlakerp2009}, the strength of the interaction depends critically on the radial distance from the Galactic center. 
Our findings (see Fig.\,\ref{posvel} and Sect.\,\ref{sec:velostruc}) are consistent with the hypothesis that LA2 and LA3 filaments probably passed through a thinner volume density portion at longer galactocentric distances than LA1.3. Using the \citet{kalberlakerp2009} model, we suggest for LA2 and LA3 a galactocentric distance of  $\geq 40$\,kpc where a density $n \leq 10^{−-4}\,{\rm cm^{-−3}}$ is expected.

\subsection{Age and size of the whole LA structure}\label{sec:symmetry}
Figure\,\ref{LA_clouds} (right panel) shows a surprisingly {\em high degree of symmetry of the \hi object distribution}. LA4 covers the same angular range in $|b_{\rm MS}|$ but is located at the opposite side of the LMC as LA1.
To scatter the \hi gas across more than $60^\circ$ on the sky, symmetrically distributed around the LMC orbit ($b_{\rm MS} = 0^\circ$), one needs a fast acceleration mechanism to overcome the gravitational attraction of the MCS and to widely distribute the gas across tens of kpc. 
Assuming a total gravitational mass of $M_{\rm MCS} = 2\cdot 10^{10}\,{\rm M_\odot}$ \citep{vandermarel02} the escape velocity is $v_{\rm LMC,esc} \simeq 280\,{\rm km\,s^{-1}}$.

The {\em high degree of symmetry in radial velocity between LA\,1 and LA\,4} implies that the whole LA is viewed nearly face on. 
Evaluating the ratio of the radial to tangential velocity vector component with the escape velocity $|v_{\rm LMC,tan}| + |v_{\rm LMC,rad}| \leq |v_{\rm LMC,esc}|$ (\citet{vandermarel02} gives $v_{\rm LMC,rad} = 84\,\pm\,7\,{\rm km\,s^{-1}}$ and $v_{\rm LMC,tan} = 281\,\pm\,41\,{\rm km\,s^{-1}}$), and we can support this visual impression quantitatively. Almost all of the total velocity vector is in the tangential component. If we assume that we watch a nearly ``face-on LA system'' with an inclination against the sky of only $13.6^\circ$, we can estimate the minimum time needed for an LA cloud to travel from the LMC body to its current location. For this, we assume a) that the linear separation between LA 4 and LA1 is less than 52\,kpc, and b) we neglect drag forces and assume a constant objects orbit velocity equal to the escape velocity. This yields $t  = 6\cdot 10^7$\,a.
According to this picture LA1.3 has a radial separation from the Galactic center of 19.3\,kpc, while LA4 lies at a galactocentric distance of 71.9\,kpc.
The LA1.3 distance estimate we derived is consistent with \citet{naomi08}, who determined $R \sim 17$\,kpc via a kinematic distance estimate. 

\subsection{Physical conditions of the LA4 clouds}\label{sec:physcond}
\subsubsection{Evaluating the LA4 ambient volume density} 

An estimated galactocentric distance of 72 kpc for LA4 corresponds to a radial distance of 74 kpc and an altitude tan(20$^\circ) \times 74$\,kpc $\simeq 26.5$\,kpc above the disk. 
Figure\,\ref{LA_individual}\,(bottom left panel) suggests that even the faint and distant LA\,4 clouds are ``decelerated'' by drag forces.
This implies that even at such large distance above the disk a sufficient high volume density exists to cause drag.
This hypothesis is supported by individual HT clouds associated with LA4. Figure\,\ref{comet1} shows the \hi 21-cm column density distribution of two clouds of LA4, where apparently both appear to be associated with a cometary shape. They appear to ``travel'' in parallel and consist of fast--moving heads and trailing tails, implying that gas is stripped off the clouds, heated up, and concentrated within the cloud's slipstream.
This gives us the chance to estimate the physical conditions at those long distances from the galaxy, so we derive from the angular extent of the head of $0.5^\circ$, with 74\,kpc distance, $N_{\rm HI} = 5.2\,\cdot10^{19}\,{\rm cm^{-2}}$, a volume density of $n_{\rm HI} = 0.03\,{\rm cm^{-3}}$. The mass of the HT clouds is about $M = 4.5\cdot 10^5\,{\rm M_\odot}$.
We estimate the kinetic temperature from the line width ($FWHM = 6.6\,{\rm km\,s^{-1}}$) to $T_{\rm kin} \leq \frac{m_{\rm H} \times (\Delta v)^2}{\rm 8\,k\,ln(2)} \simeq $\,1000\,K. This yields a gas pressure of $P = 60\,{\rm K\,cm^{-3}}$ in the certainly partly ionized gas.
To constrain the location of the HT features we assume that they move with the same velocity as the MCS. The most probable position along the line of sight would be that closest to the MCS orbit. We obtain a distance of $\sim 40 $\,kpc with a halo density of $n_{\rm Halo} = 3\cdot 10^{-3}\,{\rm cm^{-3}}$ \citep{kalb03}. This density would be just high enough to cause significant drag forces \citep{quilisandmoore01}. Figure\,\ref{comet1} shows that the $l_{\rm MS} = 23\degr $, $b_{\rm MS} = 30\degr $ cloud is positionally correlated with a bow shaped feature at $v_{\rm GSR} = 144\,{\rm km\,s^{-1}}$ (contour lines), suggestive of an interaction. Accordingly, the objects might interact with local density enhancements. An even larger bow-like emission, features are located ahead at velocities $v_{\rm GSR} = 25\,{\rm km\,s^{-1}}$.

The orientation of the projected HT velocity vector indicates its motion with respect to the ambient medium.
The resulting vector is aligned with the tail of the HT feature. For this estimate we use the mean 3-D velocity vector determined by \citet{kallivayalil2006} with $|v| = 378\,{\rm km\,s^{-1}}$.

\section{Head-tail clouds towards the LA}\label{sec:HT}

Head-tail clouds are asymmetric structures. Such asymmetries are usually determined by visual inspection, but such a description may contain personal biases. As an attempt to determine the asymmetry of these structures in an objective way, we included higher moments in our analysis. The skewness in particular appears to be an appropriate measure for head-tail clouds. A well known problem, however, is that higher moments are increasingly stronger affected by uncertainties. Unfortunately, the signal-to-noise ratio is low for most of our clouds. Skewness is then ill-defined. The column density of the tail is typically lower than that of the head. For a lower signal-to-noise ratio it gets increasingly harder to measure asymmetry since the column density of the tail may eventually be undetectable. We searched for another more robust measure of asymmetry using parameters that are less affected by noise. 
Using our catalog of LA clouds, it is feasible to systematically search for clouds that disclose a positional difference between the on the sky projected center of mass (CM) and the brightness temperature maximum (MI, max. intensity). Both parameters are not severely affected by noise but a displacement represents a measure for asymmetries.

In total we count 136 clouds with angular sizes in excess of 14.4 arcmin and with ${\rm |CM-−MI|} \geq 14.4$\,arcmin. This implies that about 30\% of our cataloged objects can be considered morphologically as HT objects.

This rate is comparable to \citet{putman11} who have found that 35\% of the HIPASS compact and semi-compact HVCs can be classified as HT clouds from their morphology. However, their 116 HT objects were selected from a total number of about 2000 HVCs on the whole southern sky. Our sample of 136 HT objects was derived from a region covering 1.6 sr, only about one quarter of the area studied by \citet{putman11}. We conclude that the LA region has a significant excess of HT structures.

The fainter one of the two HT features shown in Fig.\,\ref{comet1} has remained undetected so far.

HT HVCs are common structures of HVCs \citep{bruens2000, putman11}. Their cometary shape implies a physical interaction of the HVC gas with the dynamical different ambient medium. \citet{bruens2000} show that a linear relation exists between the peak column density of HVCs and the fraction of observed HT HVCs. This implies that the higher the peak column density, the higher the probability of identifing an HT structure.
This finding suggests that the HVC gas is compressed at the head, the gas cools more efficiently due to the higher volume density, and this leads to lower kinetic temperatures (narrower line widths) and apparently higher \hi brightness temperatures. Only towards HVC\,125+41-207 \citep{bruens2001} do we presently have evidence of this process in great detail. \citet{bruens2001} show the proposed temperature structure using a Gaussian decomposition of the Effelsberg \hi data.
On a statistical basis, the cometary shape on its own is no more than a morphological implication, nor does the linear correlation between peak column density and HT fraction directly disclose ram pressure as a physical process.

To search for physical indicators for ram pressure interaction with the ambient medium, we investigated the ratio between the brightness temperature ($T_{\rm B}$) and the kinetic temperature ($T_{\rm kin}$). 
\begin{eqnarray}
T_{\rm B} &=& T_{\rm S}\cdot (1-e^{-\tau}) \leq T_{\rm S} = T_{\rm kin} \nonumber \\
          &\leq& \frac{{\rm m_{\rm H}}\cdot \Delta v^2}{\rm 8\,k\,ln(2)}
\end{eqnarray}
   \begin{figure*}
   \centerline{
   \includegraphics[width=15cm, angle=270]{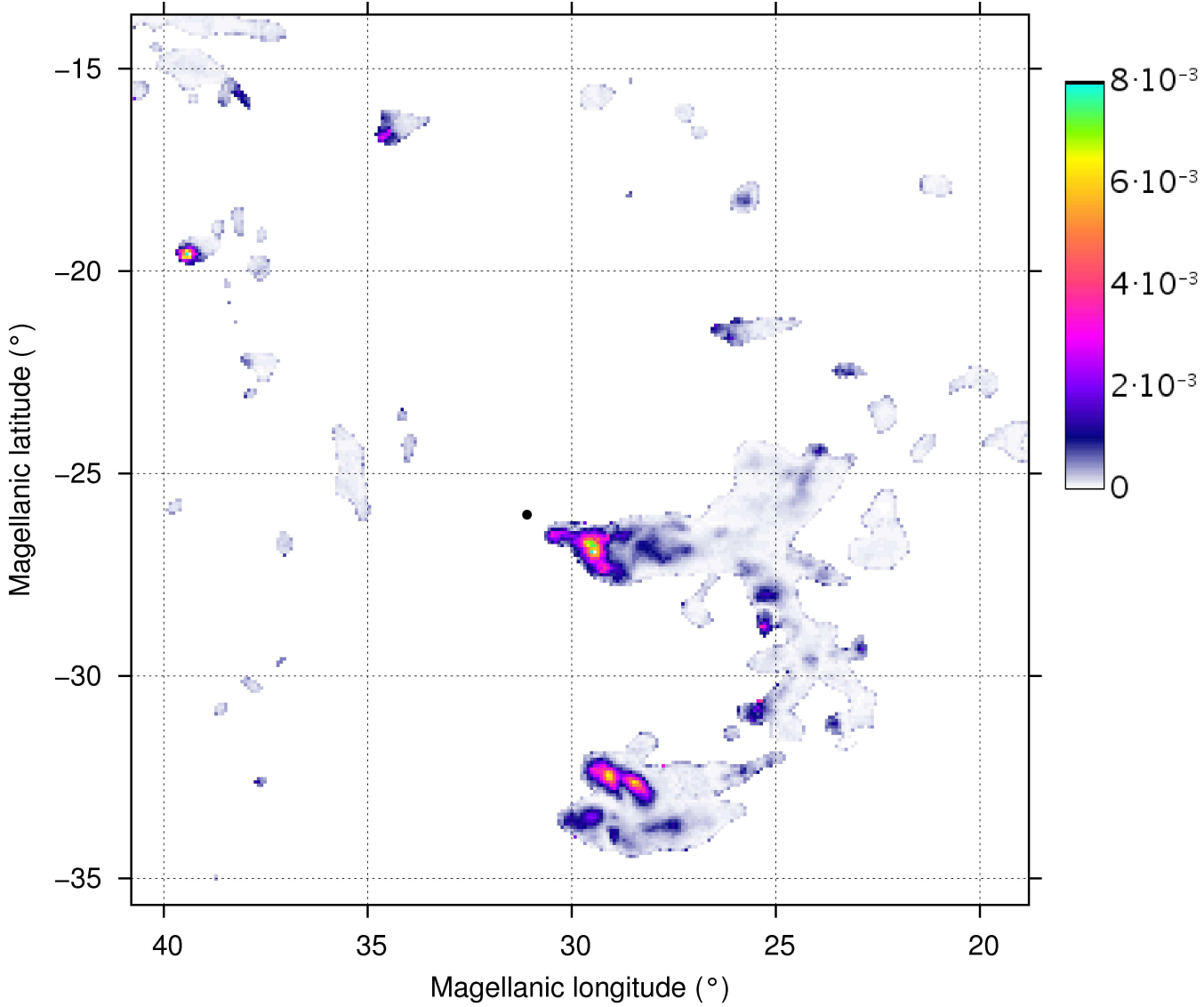}}
   \caption{Ratio of $\frac{T_{\rm B}}{T_{\rm kin}}$ as a measure of temperature and density gradients. For comparison of the linear scale we plotted the Parkes beam as a black circle in front of the LA\,1.3 head. Cooler gas portions (larger $\frac{T_{\rm B}}{T_{\rm kin}}$) and warmer (smaller $\frac{T_{\rm B}}{T_{\rm kin}}$) with a surprisingly high degree of small--scale structure. The high column density objects are associated with cool cores (heads) localized in the direction of the orbital motion (velocity gradient). 
}
              \label{fig:HTla13}
    \end{figure*}
Here $T_{\rm S}$ denotes the spin temperature, the Boltzmann population of the two energy levels, $\Delta v$ is the FWHM of the \hi emission line, and $\tau$ the opacity of the \hi gas. $T_{\rm B}$ can be considered accordingly as lower, while $T_{\rm kin}$ is an upper temperature limit of the HT cloud. The ratio of both temperatures is used as a marker to differentiate between high volume density portions of the clouds ($T_{\rm B}$ dominant) and warmer low volume density portions ($T_{\rm kin}$ dominant).
Using the LA catalog, we calculated maps of the temperature ratio across the whole LA region of interest. We focused in particular on LA\,1.3, because of the compelling morphological argumentation of \citet{naomi08}. Figure\,\ref{fig:HTla13} shows the temperature ratio. Obviously the HT cloud ($l_{\rm MS} = 27^\circ$, $b_{\rm MS} = -22^\circ$) shows a suggestive cometary shape and a ``cool'' head associated with a trailing ``warm'' tail. ``Warm'' rims are obviously common in LA\,1.3 and consistently oriented towards the Milky Way disk,
indicating physically the interaction of the LA gas with the Milky Way disk matter.

   \begin{figure*}
   \centerline{
   \includegraphics[width=17cm, angle=270]{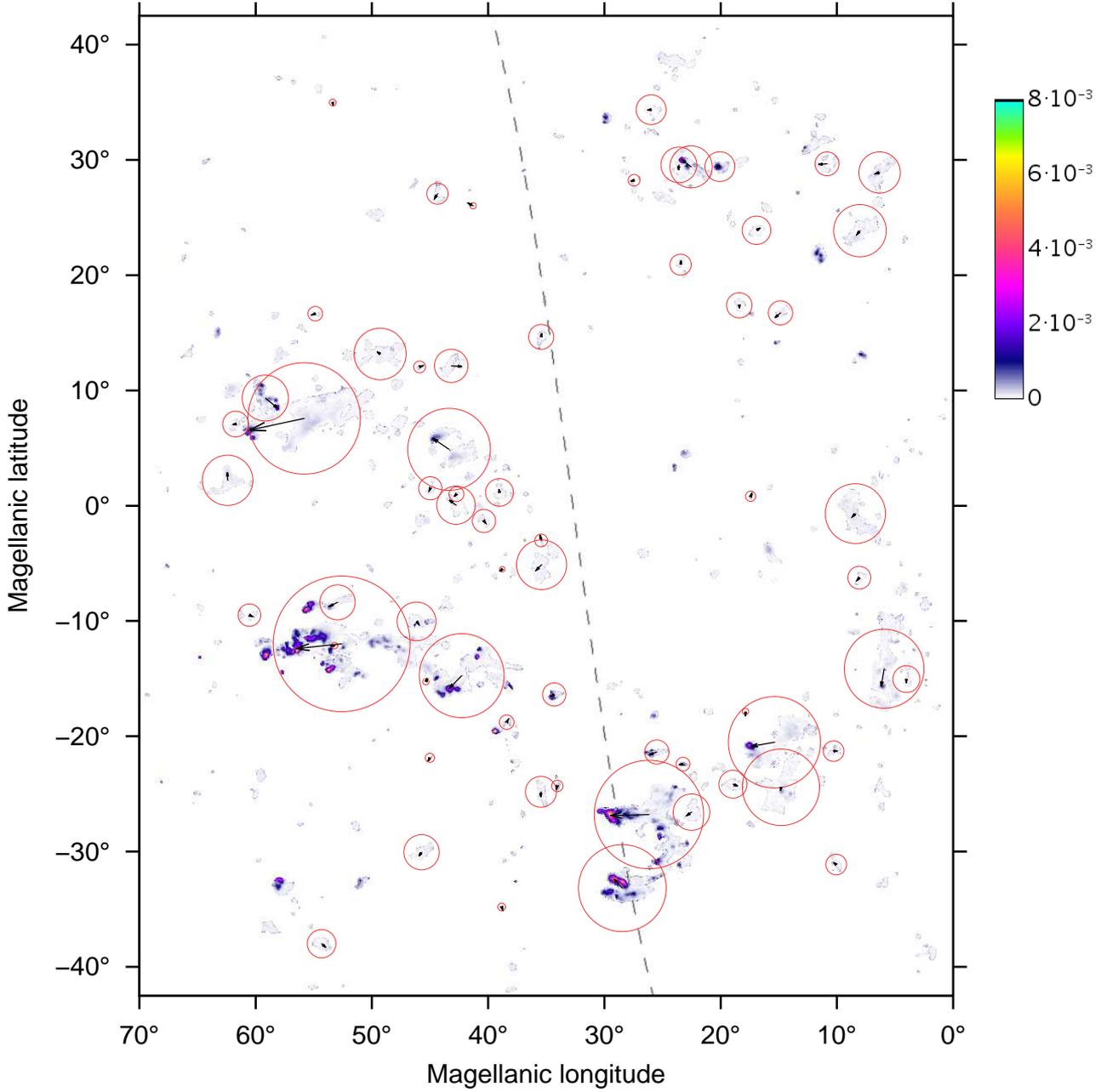}}
   \caption{Ratio of $T_{\rm B}$ and $T_{\rm kin}$ across the whole LA as shown in detail in Fig.\,\ref{fig:HTla13}. The arrows mark the orientation of the HT structures from the center of mass to the maximum brightness temperature.
}
              \label{fig:HTlatotal}
    \end{figure*}
Extending this investigation to the whole field of interest (Fig.\,\ref{fig:HTlatotal}) discloses a surprising richness of temperature structure. All HT structures have warm rims. Objects with higher masses have a cold head embedded within warm gas, followed by an even warmer trailing gas component. The low--mass objects do not show cold condensations.

Finally we studied the sky parallel component of the vector connecting CM and MI and found that the high mass objects disclose a coherent velocity pattern (Fig.\,\ref{fig:HTlatotal}). The HT velocity vectors trace the orbital motion of the LA as a whole. In combination with the derived overall velocity gradient, the velocity segregation of the object masses and the warm rims of the HT objects, we conclude that the whole LA is decelerated by drag forces towards the Milky Way disk. Both LA\,2 and LA\,3 have already passed the Milky Way disk (see Fig.\,\ref{LA_total_area}). LA\,4 fits perfectly, spatially as well as kinematically, in the whole picture and extends the LA to a huge coherent structure across the southern hemisphere. 

\section{Summary and conclusions}\label{sec:summary}
We used the GASS data \citep{naomi09, kalb10} to investigate the LA structure. The analyzed area covers $70^\circ \times 85^\circ$ degrees.
Using a sophisticated image characterization tool, it was feasible to differentiate between the diffuse Milky Way emission and the LA structures. In total
we identified 449 objects above the $4-\sigma$ threshold of the original data ($T_{\rm B} = 200\,{\rm mK}$). The applied method not only allows to determine the location of the objects, but also compiles derived information such as radial velocity, phase--space volume, column density, geometric center, and temperature weighted center. This allowed us to perform an automated statistical analysis of the LA as a whole and individual substructures down to the scales of individual clouds (HT structures). 

Next to the well studied LA1, LA2, and LA3, we discovered LA4, which is localized at high Magellanic latitudes but at comparable longitudes to LA1. This symmetry but also the associated velocity gradient, suggests that this feature belongs to the LA complex. LA4 on average contains less extended clouds. In Sect. \ref{sec:velostruc} we argued for a distance between 40 and 70 kpc, which implies that these clouds could also have a lower average mass.

Averaging across the whole LA1 to LA4 cataloged clouds, we derived a clear negative velocity gradient (deceleration) with Magellanic stream longitude. Assuming $D = 50$\,kpc, the mass weighted linear correlation yields $v_{\rm GSR} = 84.2\,{\rm km\,s^{-1}}$ as axis intercept. This value is identical within its uncertainties with the bulk velocity of the LMC derived by \citet{vandermarel02} from the carbon star ensemble. According to this we support the hypothesis of \citet{nidever08} that the LA as a whole has its origin in the LMC.

The mass of the LA is about a factor of 50\% higher than previously assumed (\citet{bruens05} at a distance of 50\,kpc, \citet{cioni00}). The main reason for this increase is that our survey covered a significantly larger region than is available for previous, more specialized investigations. Some objects, blended with Milky Way emission features, may be missing in our analysis. The LA mass of $38 \cdot 10^{6}\,{\rm M_\odot}$ derived here is probably a lower limit only.

The velocity structure of individual LA filaments can be interpreted as prototypical for deceleration by drag forces. LA1.3 shows the strongest velocity gradient, consistent with the proposal of \citet{naomi08}. All coherent structures of the LA but LA1.1 and LA1.2 show a systematic deceleration towards the Galactic plane. We find an obvious difference in the mass vs. velocity distribution for LA1 and LA4 in comparison to LA2 and LA3. The last two show that the higher the mass, the higher the $v_{\rm GSR}$ velocity. In the framework of drag models, these high--mass compact objects keep their momentum, while the low--mass, more diffuse clouds are significantly decelerated.

The apparent symmetry of the LA structure is a surprising finding. The whole structure covers nearly $65^\circ$ across the whole sky. At a distance of 50\,kpc, this corresponds to a linear scale of 52.5\,kpc. Combining the radial velocity information of the \hi data with the proper motion \citep{vandermarel02}, we estimate an inclination of only $13.6^\circ$ against the plane of the sky. With the escape velocity for the LMC we can calculate that the LA structures are at least $6\cdot 10^7$\,a old.
This nearly face-on geometry of the LA localized LA1.3 at a galactocentric distance of $R\sim 19.3$\,kpc and LA4 at $R\sim 71.9$\,kpc.

We find pronounced HT structures associated with LA4. Using a distance of 74\,kpc, an estimate on the kinetic temperature by the \hi line width, we estimate a pressure of $P = 60\,{\rm K\,cm^{-3}}$. According to \citet{kalb03} the ambient volume density with $n \simeq 10^{-3}\,{\rm cm^{-3}}$ is high enough to cause drag.

The analysis of the ratio between $T_{\rm B}$ and $T_{\rm kin}$ implies that the gas of the objects is compressed towards the leading edge of the objects. Warm rims are at the head of the HT structures, suggesting gas compression and eventually an enhanced cooling. All massive LA objects show HT structures. The sky projected vector component between the location of their centers of mass relative to the maximum \hi line intensities are all oriented parallel to the orbital motion of the MCS. In combination with the velocity segregation, the overall velocity gradient and the cool leading rims of the HT objects, we conclude that the whole LA is a coherent structure decelerated by drag. The newly identified LA\,4 fits in the whole structure and physical interpretation perfectly. 

The emerging picture of a four-arm structure, with two symmetric arms approaching the Milky Way disk (LA1 and LA4) and two arms that have already passed the disk (LA2 and LA3) provides new constraints for simulations of the orbital history of the Magellanic Clouds. \citet{DiazBekki2011} present a new tidal model in which, for the first time, structures resemble a clear bifurcation of the LA. Concerning LA4, we would like to note that the existence of this feature was predicted by \citet{Besla2010}, but their simulations fail to recover LA1. Hydrodynamic processes such as ram pressure are needed to shape the Magellanic Stream. 

\begin{acknowledgements}
MSV thanks the \hi group of the Argelander-Institut f\"ur Astronomy for a stipend during the preparation phase of the manuscript. The JK and PMWK acknowledge the financial support from the Deutsche Forschungsgemeinschaft (DFG) under the reseach grants KE757/7-1, KE757/7-2, and KE757/9-1. The authors like to thank an anonymous referee for very helpful comments and suggestions.
\end{acknowledgements}

\bibliographystyle{aa}
\bibliography{references}

\begin{longtable}{lllrr}\\
\caption{\label{tab:LA_Objects} Excerpt of the LA catalog of objects available at the CDS. The table comprises several parameters that describe the 449 LA objects discussed.}\\
\hline
Units  &  Label &  Explanations & Object 1 & Object 2 \\
\hline
-	&	Seq	&	  Sequential number	&	1	&	2	\\
-	&	ID	&	  Internal ID number	&	258	&	259	\\
-	&	Name	&	  High-velocity cloud (HVC) name (1)	&	HVC 291.7+21.5+111.1	&	HVC 293.4+37.1+086.4	\\
${\rm deg^2\,km\,s^{-1}}$	&	PhaVol	&	  Phase space volume in ${\rm deg^2\,km\,s^{-1}}$	&	0.11	&	1.19	\\
solMass	&	MassMsol	&	  Mass at 50\,kpc in solar masses	&	323.39	&	3848.43	\\
sr	&	SolAng	&	  Solid angle	&	2.07E-05	&	5.23E-05	\\
pc$^2$	&	Area	&	  Area at 50\,kpc from solid angle	&	51795.3	&	130910.9	\\
deg	&	SphDiam	&	  Spherical diameter from solid angle	&	0.29	&	0.46	\\
pc	&	SphDiampc	&	  Spherical diameter at 50\,kpc from area	&	256.8	&	408.2	\\
cm$^{-2}$	&	PeakNHI	&	  Peak column density	&	1.53E+018	&	7.81E+018	\\
cm$^{-2}$	&	MeanNHI	&	  Mean column density	&	7.79E+017	&	3.66E+018	\\
cm$^{-3}$	&	Density	&	  Mean cloud density at 50\,kpc	&	9.80E-04	&	3.00E-03	\\
K\,cm$^{-3}$	&	Press	&	  Mean cloud pressure at 50\,kpc in K\,cm$^{-3}$	&	0.05	&	5.23	\\
-	&	LocMax	&	  Local maxima $>$\,0.24\,deg and $>$\,2.4\,km\,s$^{-1}$	&	3	&	6	\\
-	&	RegMax	&	  Regional maxima $>$\,0.56\,deg and $>$\,5.6\,km\,s$^{-1}$	&	1	&	2	\\
deg	&	MSLON	&	  CM Magellanic stream longitude	&	53.04	&	69.30	\\
deg	&	MSLAT	&	  CM Magellanic stream latitude	&	-14.93	&	-16.40	\\
deg	&	GLON	&	  CM Galactic longitude	&	291.65	&	293.42	\\
deg	&	GLAT	&	  CM Galactic latitude	&	21.48	&	37.13	\\
deg	&	RAdeg	&	  CM right ascension (J2000)	&	179.08	&	184.49	\\
deg	&	DEdeg	&	  CM declination (J2000)	&	-40.17	&	-25.12	\\
km\,s$^{-1}$	&	HRV	&	  CM heliocentric radial velocity	&	115.81	&	88.05	\\
km\,s$^{-1}$	&	LRV	&	  CM local standard of rest velocity	&	111.08	&	86.35	\\
km\,s$^{-1}$	&	GRV	&	  CM Galactic standard of rest velocity	&	-79.17	&	-74.58	\\
km\,s$^{-1}$	&	LGRV	&	  CM local group standard of rest velocity	&	-147.88	&	-144.62	\\
km\,s$^{-1}$	&	FWHM	&	  CM velocity FWHM	&	1.64	&	11.54	\\
-	&	xxVar	&	  CM xxVariance (2)	&	0.17	&	1.03	\\
-	&	yyVar	&	  CM yyVariance (2)	&	1.16	&	1.20	\\
-	&	xyVar	&	  CM xyVariance (2)	&	0.19	&	-0.19	\\
-	&	xSkew	&	  CM xSkewness (2)	&	0.57	&	-0.00	\\
-	&	ySkew	&	  CM ySkewness (2)	&	-0.23	&	-0.08	\\
-	&	xKurt	&	  CM xKurtosis (2)	&	-0.97	&	-0.87	\\
-	&	yKurt	&	  CM yKurtosis (2)	&	-1.09	&	-1.01	\\
deg	&	PA	&	  CM position angle of ellipse to north (3)	&	150.6	&	38.9	\\
-	&	Ecc	&	  CM eccentricity of ellipse (3)	&	0.94	&	0.62	\\
deg	&	MinDiam	&	  CM minor diameter of ellipse (3)	&	0.16	&	0.41	\\
deg	&	MajDiam	&	  CM major diameter of ellipse (3)	&	0.46	&	0.53	\\
deg	&	PTBLON	&	  Peak T$\rm_B$ Magellanic stream longitude	&	53.00	&	69.24	\\
deg	&	PTBLAT	&	  Peak T$\rm_B$ Magellanic stream latitude	&	-14.96	&	-16.40	\\
km\,s$^{-1}$	&	PTBGRV	&	  Peak T$\rm_B$ GSR velocity	&	-78.32	&	-73.38	\\
deg	&	PTBAng	&	  Angle between CM - peak T$\rm_B$ line and north	&	119.5	&	90.0	\\
K	&	PeakTB	&	  Peak brightness temperature (T$\rm_B$)	&	0.32	&	0.45	\\
km\,s$^{-1}$	&	PTBFWHM	&	  Peak T$\rm_B$ velocity FWHM	&	1.64	&	9.07	\\
K	&	TDopp	&	  Peak T$\rm_B$ Doppler temperature at 50\,kpc	&	59.3	&	1796.3	\\
pc	&	SizeLWRel	&	  Peak T$\rm_B$ Size-linewidth relation	&	1.6	&	287.7	\\
deg	&	CenLON	&	  Centroid Magellanic stream longitude (4)	&	53.04	&	69.30	\\
deg	&	CenLAT	&	  Centroid Magellanic stream latitude (4)	&	-14.93	&	-16.40	\\
km\,s$^{-1}$	&	CenGRV	&	  Centroid GSR velocity (4)	&	-79.18	&	-74.57	\\
deg	&	SizeLON	&	  Size along Magellanic stream longitude	&	0.23	&	0.61	\\
deg	&	SizeLAT	&	  Size along Magellanic stream latitude	&	0.48	&	0.40	\\
km\,s$^{-1}$	&	SizeVel	&	  Size along radial velocity	&	3.29	&	12.36	\\
\hline
\end{longtable}

\end{document}